\documentclass[twocolumn]{aastex62}
\usepackage{amsmath}
\usepackage{amssymb}
\usepackage{array,multirow}
\usepackage{comment}
\usepackage{enumerate}
\usepackage{bm}
\usepackage{ulem}

\usepackage[flushleft]{threeparttable}

\usepackage{xcolor}
\definecolor{OliveGreen}{rgb}{0,0.6,0}
\definecolor{Salmon}{rgb}{1, 0.549, 0.412}

\newcommand{\degree}{\ifmmode {^{\circ}\ }\else$^{\circ}$\fi}

\newcommand{\Msun}{\ifmmode {M_{\odot}}\else${M_{\odot}}$\fi}
\newcommand{\Rsun}{\ifmmode {R_{\odot}}\else${R_{\odot}}$\fi}
\newcommand{\Porb}{\ifmmode {P_{\rm orb}}\else${P_{\rm orb}}$\fi}
\newcommand{\Pspin}{\ifmmode {P_{\rm spin}}\else${P_{\rm spin}}$\fi}
\newcommand{\Pdot}{\ifmmode {\dot{P}_{\rm spin}}\else${\dot{P}_{\rm spin}}$\fi}
\newcommand{\age}{\ifmmode {\tau_{\rm c}}\else${\tau_{\rm c}}$\fi}

\newcommand{\katie}[1]{\textbf{\color{Salmon} #1}}

\submitjournal{ApJL}

\shorttitle{DNSs in LISA}
\shortauthors{Andrews, J. J. et al. }

\begin{document}

\title{LISA and the Existence of a Fast-Merging Double Neutron Star Formation Channel}

\author[0000-0001-5261-3923]{Jeff J. Andrews}
\affiliation{Niels Bohr Institute, University of Copenhagen, Blegdamsvej 17, 2100 Copenhagen, Denmark}
\email{jeff.andrews@nbi.ku.dk}

\author[0000-0001-5228-6598]{Katelyn Breivik}
\affiliation{Canadian Institute for Theoretical Astrophysics, University of Toronto, 60 St. George Street, Toronto, Ontario, M5S 1A7, Canada}

\author[0000-0002-1128-3662]{Chris Pankow}
\affiliation{Center for Interdisciplinary Exploration and Research in Astrophysics (CIERA) and Department of Physics and Astronomy, Northwestern University, 2145 Sheridan Road, Evanston, IL 60208, USA}

\author[0000-0002-1271-6247]{Daniel J. D'Orazio}
\affiliation{Center for Astrophysics, Harvard \& Smithsonian, 60 Garden Street, Cambridge, MA 02138, USA}

\author[0000-0002-1827-7011]{Mohammadtaher Safarzadeh}
\affiliation{Department of Astronomy and Astrophysics, University of California, Santa Cruz, CA 95064}

\begin{abstract}
Using a Milky Way double neutron star (DNS) merger rate of 210 Myr$^{-1}$, as derived by the Laser Interferometer Gravitational-Wave Observatory (LIGO), we demonstrate that the Laser Interferometer Space Antenna (LISA) will detect on average 240 (330) DNSs within the Milky Way for a 4-year (8-year) mission with a signal-to-noise ratio greater than 7. Even adopting a more pessimistic rate of 42 Myr$^{-1}$, as derived by the population of Galactic DNSs, we find a significant detection of 46 (65) Milky Way DNSs. These DNSs can be leveraged to constrain formation scenarios.
%
In particular, traditional NS-discovery methods using radio telescopes are unable to detect DNSs with \Porb\ $\lesssim$1 hour (merger times $\lesssim$10 Myr). If a fast-merging channel exists that forms DNSs at these short orbital periods, LISA affords, perhaps, the only opportunity to observationally characterize these systems; we show that toy models for possible formation scenarios leave unique imprints on DNS orbital eccentricities, which may be measured by LISA for values as small as $\sim$10$^{-2}$.
\end{abstract}


\section{Introduction}
\label{sec:intro}

Current population synthesis models predict a double neutron star (DNS) merger rate within the Milky Way of $\approx$20-40 Myr$^{-1}$ \citep{vigna-gomez18, kruckow18, chruslinska18, mapelli18}. Alternatively, the DNS merger rate can be calculated from the merger times of the known DNSs in the Milky Way field, accounting for survey selection effects \citep{phinney91b, kim03}. 
The latest application of this method, using 17 DNSs in the Milky Way field, finds a rate of 42$^{+30}_{-14}$ Myr$^{-1}$ 
although this estimate is sensitive to pulsar luminosities, lifetimes, assumptions about the contribution from elliptical galaxies, and beaming correction factors \citep{pol19}. 
In comparison, a volumetric DNS merger rate of 920$^{+2220}_{-790}$ Gpc$^{-3}$ yr$^{-1}$ \citep[which translates into a Milky Way rate of $\approx$210 Myr$^{-1}$;][]{kopparapu08}, can be derived from the second Laser Interferometer Gravitational-Wave Observatory (LIGO)/Virgo observing run \citep[O2;][]{ligo_detection, ligo_GWTC1}. 
While the relatively large errors make the two observational rate estimates consistent, with values dependent on various assumptions about NS population properties, the rate derived from LIGO \citep{ligo_GWTC1} is nevertheless somewhat larger than the analogous rate derived from the DNS Milky Way field population.

One possible origin of this difference could arise from the methods by which DNSs are detected within pulsar surveys. The known DNSs in the Milky Way have orbital periods ($\Porb$) ranging from as large as 45 days \citep[J1930$-$1852;][]{swiggum15} to as small as 1.88 hours \citep[J1946$+$2052;][]{stovall18}. Even shorter period binaries may exist, but are extremely challenging to find for two reasons. First, they quickly merge due to general relativistic (GR) orbital decay; since $t_{\rm merge}\sim \Porb^{8/3}$, it is much more likely to observe DNSs which form with longer orbital periods. Second, Doppler smearing reduces the sensitivity of pulsar surveys to binaries with orbital periods $\lesssim$ hours \citep{bagchi13}. The detection of such short-period binaries typically requires acceleration searches, which are both technically challenging and computationally expensive \citep[see e.g.,][]{ng15}. The difficulty of identifying DNSs at shorter periods than a few hours will cause DNS merger rate estimates based on Milky Way populations, such as those by \citet{pol19}, to be systematically underestimated by an amount that depends on the number of systems formed at these short orbital periods; while these estimates account for detection biases for the observed systems, they cannot account for systems that are formed with \Porb\ so short that selection effects make them effectively un-observable.

\begin{figure}
	\includegraphics[width=0.99\columnwidth]{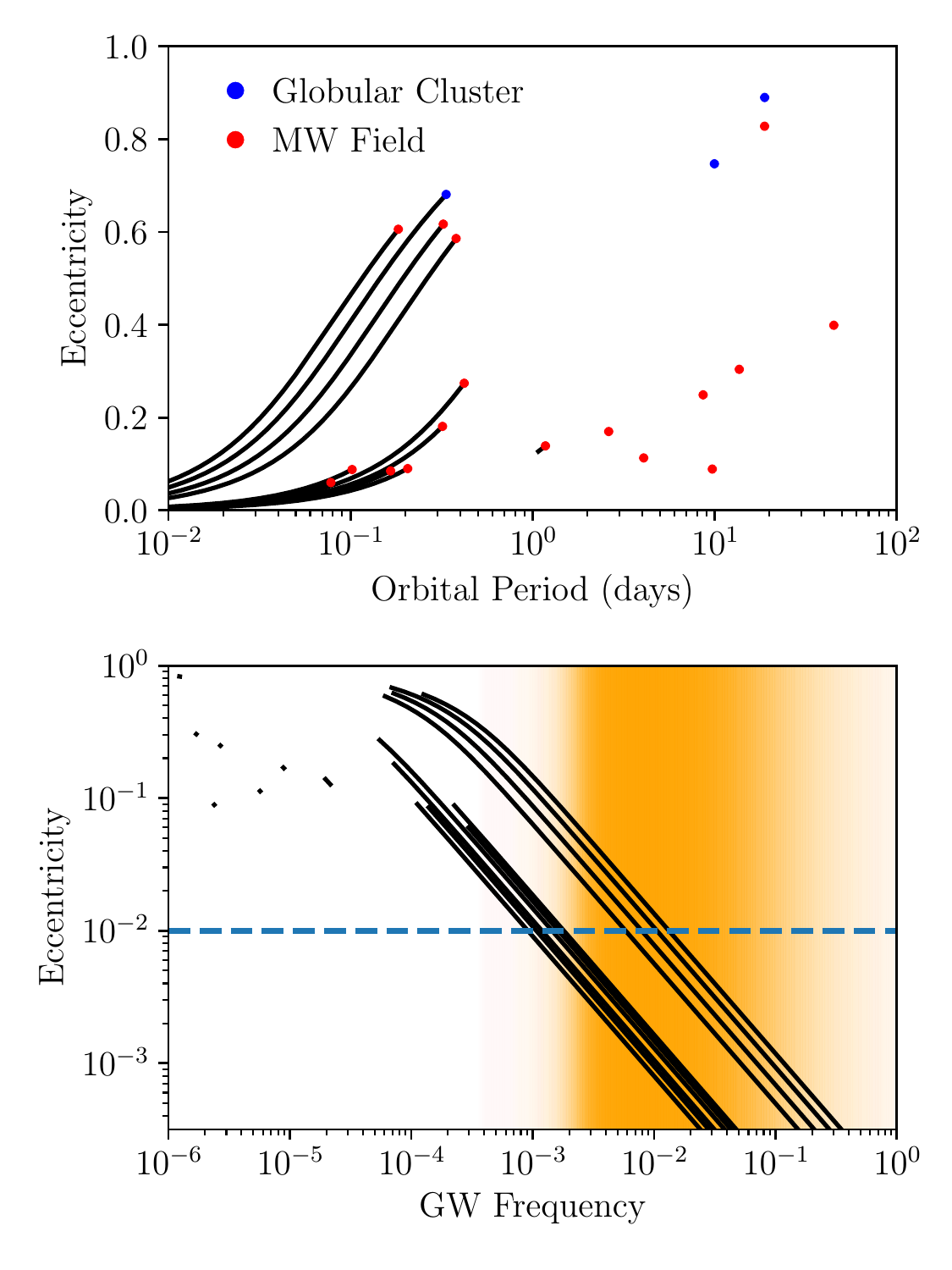}
    \caption{{\bf Top panel:} The sample of 20 DNSs in the Milky Way (red points). Uncertainties on the measured \Porb\ and $e$ are smaller than the data points. Black lines indicate the evolution of these orbits as these systems circularize and decay due to gravitational wave radiation in a Hubble time. {\bf Bottom panel:} These DNSs retain residual eccentricities ($e\gtrsim10^{-3}$) as they evolve through the LISA band (the relative sensitivity of LISA is represented by the orange background). Pre-empting our results in Section \ref{sec:eccentricity}, LISA can measure binary eccentricities as small as $\sim$10$^{-2}$ (blue, dashed line) for typical DNSs. Therefore, LISA will measure the eccentricities of many DNSs (which typically have $f_{\rm GW}\approx10^{-2.75}$ Hz; see Figure \ref{fig:lisa_sensitivity}).}
    \label{fig:DNS_evolution}
\end{figure}

While radio observations are sensitive to DNSs with $P_{\rm orb}$\,$\gtrsim$\,hrs and LIGO/Virgo detects DNSs at merger, the Laser Interferometer Space Antenna \citep[LISA;][]{lisa_paper} is sensitive to binaries with $P_{\rm orb}\sim$minutes, bridging the gap between these two regimes. We use the equations from \citet{peters64} to show in the top panel of Figure \ref{fig:DNS_evolution} how the orbits of the 20 known DNSs in the Milky Way \citep[17 in the field, 3 in globular clusters; see][and references therein for a list]{ridolfi19, andrews19b} will evolve over the next 10 Gyr as GR causes them to circularize and decay. Depending on eccentricity, DNSs born with orbital periods longer than $\approx$18 hours take longer than the age of the Universe to merge due to GR.  The bottom panel of Figure \ref{fig:DNS_evolution} shows the evolution of the orbital eccentricity with the gravitational wave frequency ($f_{\rm GW} = 2/\Porb$), of these same 20 systems. 

If sufficiently close to the Solar System, binaries with 10$^{-4}$ Hz $<$ $f_{\rm GW}$ $<$ 1 Hz are detectable by LISA. Although GR circularizes orbits as they inspiral, DNSs formed in eccentric orbits will still maintain a residual eccentricity as they evolve through the LISA band.
Recent estimates suggest that LISA may be able to measure orbital eccentricities in more massive black hole binaries down to a level of 10$^{-3}$ \citep{nishizawa16}. 
Pre-empting our quantitative results in Section \ref{sec:results}, we find that LISA will have have only a slightly degraded precision for DNS binaries, measuring the eccentricities of typical systems as small as a few 10$^{-2}$ at $f_{\rm GW}\approx10^{-2.5}$ Hz. This is in agreement with recent results by \citet{lau19} who also study the detectability of DNSs with LISA. Comparison with the tracks in the bottom panel of Figure \ref{fig:DNS_evolution} shows this precision is sufficient to measure DNS eccentricities as they evolve through the LISA band. Thus, eccentricity measurements by LISA \citep{lau19} may be used to inform DNS formation scenarios in a similar fashion to binary black holes \citep{breivik2016, nishizawa17, SamsingDOrazio:2018, DOrazioSamsing:2018}.


If no DNSs are formed with orbital periods $\lesssim$1 hour, the distributions of DNS inspirals detected by LISA ought to match expectations from the tracks of the observed Milky Way DNS population shown in the bottom panel of Figure \ref{fig:DNS_evolution}. 
However, if an evolutionary channel exists that forms significant numbers of DNSs with \Porb$\lesssim$1 hour (i.e. a `fast-merging' channel), 
such systems will both increase the DNS merger rate and produce distinct tracks in the $f_{\rm GW}-e$ plane. In this work, we demonstrate that LISA may be able to measure both effects. We first calculate the number of DNSs detectable by LISA using the up-to-date LISA sensitivity curve and the current DNS merger rate in Section \ref{sec:LISA}. We then produce toy models for different fast-merging evolutionary channels in Section \ref{sec:results} and discuss LISA's ability to discern between these channels by measuring their orbital eccentricities.
We finish in Section \ref{sec:conclusions} with a discussion of our results and our conclusions.

\begin{figure*}
	\includegraphics[width=0.97\textwidth]{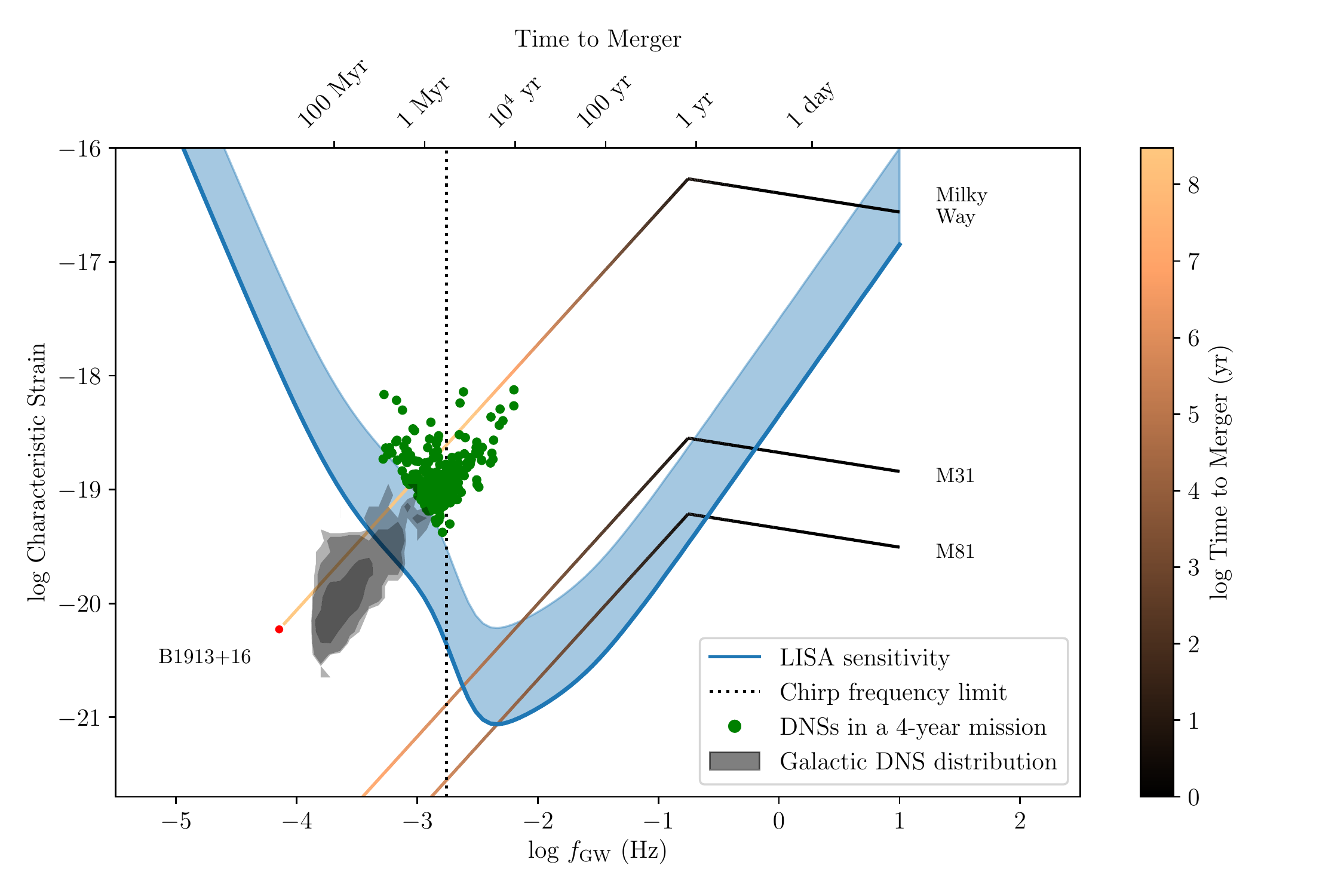}
    \caption{We compare the distribution of DNSs expected within the Milky Way (grey contours; see Section \ref{sec:MW_pop} for details) against the LISA's characteristic strain (blue line). The blue contour extends the sensitivity curve to approximately account for an SNR$=7$ detection. For one random Milky Way realization, green points indicate the DNSs with SNR$>$7, calculated using a DNS merger rate of 210 Myr$^{-1}$. We indicate the Hulse-Taylor binary, B1913$+$16, as a red point, and follow its evolution by straight lines as the system merges due to GR. The line's color indicates the time to merger. Placing B1913$+$16 at larger distances corresponding to M31 or M81 shows LISA may be able to detect a few DNSs outside of the Milky Way \citep{seto19}. Note that all binaries are plotted as though they are circular. We properly account for eccentricity when calculating the SNR for a LISA detection (see Section \ref{sec:SNR}).}
    \label{fig:lisa_sensitivity}
\end{figure*}

\section{Detecting DNS with LISA}
\label{sec:LISA}


Observatories such as LISA detect gravitational waves by measuring the slight perturbations they cause to space-time as they propagate through the detector. For a circular binary at a distance, $d$, and with a chirp mass, $\mathcal{M}_c$ (for two stars with masses $M_1$ and $M_2$, $\mathcal{M}_c=M_1^{3/5}M_2^{3/5}(M_1+M_2)^{-1/5}$), the amplitude of this strain can be determined as a function of $f_{\rm GW}$:
\begin{equation}
    h(f_{\rm GW}) = \frac{8}{\sqrt{5}} (\pi f_{\rm GW})^{2/3} \frac{(\mathcal{G}\mathcal{M}_c)^{5/3}}{c^4} \frac{1}{d},
    \label{eq:h_c}
\end{equation}
where $\mathcal{G}$ is the Gravitational constant and $c$ is the speed of light. The coefficient is set to account for averaging of the wave polarization, sky position, and DNS orientation. 
Keeping $f_{\rm GW}$ and $h(f_{\rm GW})$ constant, we find that the observable volume for a gravitational wave signal scales with $\mathcal{M}_c^5$. Despite the expectation that DNSs ought to be more than an order of magnitude more common in the Universe than black hole binaries \citep{ligo_GWTC1}, the different horizon distances between the two types of binaries will make DNSs significantly rarer within LISA than their more massive black hole binary counterparts. 

To quantitatively determine the detectability of DNSs by LISA, we use the LISA sensitivity curve described by \citet{cornish17} and \citet{Robson19}. The blue line in Figure \ref{fig:lisa_sensitivity}, which includes both the intrinsic detector sensitivity as well as the contribution from the double white dwarf foreground \citep{korol17}, denotes the sensitivity curve in terms of the characteristic strain $(f_{\rm GW} S_{\rm LISA})^{1/2}$ \citep{Robson19} as a function of $f_{\rm GW}$. Below, in turn, we first describe peculiarities of how an individual binary is detected by LISA using the Hulse-Taylor binary as an example. We then calculate the expectation from a Milky Way population as well as populations from the nearby M31 and M81 galaxy groups.


\subsection{Calculating the SNR for a LISA Detection}
\label{sec:SNR}

The signal-to-noise ratio (SNR) of a LISA detection can be calculated from the binary's strain amplitude $h(f_{\rm GW})$, LISA's noise power spectral density $S_{\rm LISA}(f_{\rm GW})$\footnote{$S_{\rm LISA}$ is taken from \citet{Robson19} who denote this function as $S_n$. We opt to adopt a different subscript to avoid confusion between their $n$ for ``noise" and the index specifying the orbital harmonic.} and LISA's lifetime $T_{\rm LISA}$ \citep{Robson19}:
\begin{equation}
    {\rm SNR}^2 = \frac{h^2(f_{\rm GW}) T_{\rm LISA}}{S_{\rm LISA}(f_{\rm GW})}.
    \label{eq:SNR_circular}
\end{equation}
Most definitions of LISA's SNR include an integral over $f_{\rm GW}$, since gravitational waves cause a binary's orbit to decay over LISA's lifetime. However, for binaries that evolve slowly, such as the DNSs that LISA will detect, the integral can be accurately approximated by Equation \ref{eq:SNR_circular} \citep{Robson19}, where we have absorbed various coefficients and prefactors into the definition of $h(f_{\rm GW})$ in Equation \ref{eq:h_c}.

In stark contrast with circular binaries, eccentric binaries emit gravitational waves at multiple harmonics of the orbital frequency. Therefore, for a binary with an eccentricity $e$, the SNR can be calculated from the quadrature sum of the SNRs for each $f_n$ harmonic of the orbital frequency \citep[see e.g.,][]{DOrazioSamsing:2018, kremer18a}
\begin{equation}
 {\rm SNR}^2 \approx \sum_{n=1}^{\infty} \frac{h_n^2(f_n) T_{\rm LISA}}{S_{\rm LISA}(f_n)}.
\end{equation}
The strain amplitude $h_n(f_n)$ can be calculated as a function of $e$ and orbital frequency harmonic $n$:
\begin{equation}
    h_n(f_n) = \frac{8}{\sqrt{5}} \left(\frac{2}{n}\right)^{5/3} \frac{(\pi f_n)^{2/3} (\mathcal{G}\mathcal{M})^{5/3}}{c^4d} \sqrt{g(n,e)},
\end{equation}
where $g(n,e)$ provides the relative amplitude at each harmonic \citep{peters63}.

\subsection{Measuring the Eccentricity}
\label{sec:eccentricity}

For LISA to measure the eccentricity of a binary, the power of at least two harmonics must be measured; the harmonics at 2/\Porb\ and 3/\Porb\ are strongest for binaries with $e\lesssim0.3$ \citep{seto01}. Since the relative ratios of the amplitudes of these two peaks provide the eccentricity measurement, following \citet{seto16}, we can use error propagation to estimate the precision of an eccentricity measurement:
\begin{equation}
    \left(\Delta e\right)^2 \approx \frac{e^2}{\left(1/{\rm SNR}_2\right)^2 + \left(1/{\rm SNR}_3\right)^2}, \label{eq:de}
\end{equation}
where SNR$_2$ and SNR$_3$ are the signal-to-noise ratios for the 2/\Porb\ and 3/\Porb\ harmonics, respectively. For $e\lesssim0.3$, the ratio of the amplitudes of the two harmonics scale with $(9/4)e$ \citep{seto16}. The SNR of a particular harmonic depends both on the amplitude of the GW signal as well as LISA's sensitivity at that frequency. Using a spectral index, $\alpha$, to describe LISA's sensitivity as a function of frequency, we can determine the relative SNR for the two harmonics
\begin{equation}
    \frac{{\rm SNR}_3}{{\rm SNR}_2} = \left( \frac{2}{3} \right)^{-\alpha-2} e. \label{eq:SNR_ratio}
\end{equation}
In the limit that eccentricities are small (so that SNR$_2>>$ SNR$_3$ and therefore the $1/\text{SNR}_2$ term can be ignored in Equation \ref{eq:de}), we find:
\begin{equation}
    \Delta e \approx \left( \frac{2}{3} \right)^{\alpha+2} \left( \frac{1}{{\rm SNR}_2} \right) . \label{eq:delta_e}
\end{equation}
For a DNS with 10$^{-2.5}$ Hz $<f_{\rm GW} < 10^{-3}$ Hz, the DWD foreground causes $\alpha=3$. Therefore, a DNS with SNR$_2=10$ has $\Delta e \approx 0.01$. For more eccentric binaries, higher order harmonics become relevant and Equation \ref{eq:delta_e} is no longer applicable. However, with multiple detected harmonics the measurement precision on $e$ will only improve.

\subsection{The Hulse-Taylor Binary as an Example}

In Figure \ref{fig:lisa_sensitivity} we show the characteristic strain ($h_c = h(f_{\rm GW})f_{\rm GW}^{1/2} T_{\rm LISA}^{1/2}$) of the Hulse-Taylor binary (red point) with its current orbital parameters at its distance to the Sun. Note that the characteristic strains shown in Figure \ref{fig:lisa_sensitivity} are calculated using Equation \ref{eq:h_c} (for plotting purposes only, we assume binaries are circular). Over the course of the next $\approx$200 Myr, the system will inspiral due to GR orbital decay. We show the path the Hulse-Taylor binary will take in $f_{\rm GW}$-$h_c$ space (assuming a constant distance) in Figure \ref{fig:lisa_sensitivity} by a line, whose color is dictated by the time remaining until the system merges. 

Scaling this curve to the distances to the nearby galaxies M31 and M81 shows that, in principle, DNSs in these galaxies produce a gravitational wave signal in the LISA band above the background. However, typically a limiting signal-to-noise ratio (SNR) of 7 (indicated approximately by the upper limit of the blue contour in Figure \ref{fig:lisa_sensitivity}) is used to determine if LISA is likely to detect a particular system \citep[e.g.,][]{korol18}. A more careful analysis using the SNR calculation provided in Section \ref{sec:SNR} is required to determine the actual number of detectable DNSs expected within the Milky Way and other nearby galaxies.



\subsection{The Milky Way Population of DNSs}
\label{sec:MW_pop}

We first focus on the Milky Way. Using the DNS merger rate estimate of 210 Myr$^{-1}$, as derived by LIGO \citep{ligo_GWTC1}, we can estimate where the DNSs would lie in $h_c$-$f_{\rm GW}$ space for a random realization of the Milky Way, assuming a steady-state merger rate. Whereas \citet{kyutoku19} have recently calculated the number of detectable DNSs in the Galaxy by LISA using an analytic approximation, we use a Monte Carlo method which allows us to account for the spatial distribution of DNSs throughout the Milky Way. \citet{lau19} also use a Monte Carlo method to generate LISA detection predictions for a population of Milky Way DNSs; however these authors base their results on the population synthesis results from \citet{vigna-gomez18}, who find a significantly lower Milky Way DNS merger rate of 33 Myr${-1}$.

We first generate 2100 random merger times (these times represent how long before an individual binary will merge), chosen from a uniform distribution over the past 10 Myr (210 Myr$^{-1}$ $\times$ 10 Myr = 2100). Extending this procedure to longer merger times is unnecessary as LISA is only sensitive to DNSs that will merge in the next $\sim$10 Myr (see Figure \ref{fig:lisa_sensitivity}). 
We initialize each of these systems with very small \Porb\ and $e$, with values equal to those that B1913$+$16 will take immediately prior to merger ($P_{\rm orb}=0.2$ s, $e=5\times10^{-6}$), then integrate their orbital evolution backwards in time for each of the randomly chosen times. The resulting set of backwards-evolved $(\Porb, e)$ values represents a Milky Way population of DNSs that produces an average merger rate of 210 Myr$^{-1}$.

We then place each simulated DNS in a random position in the Milky Way, following the Galactic model from \citet{nelemans01}:
\begin{equation}
    P(r, z) \sim e^{-R/L}\ {\rm sech}\left(z/\beta \right)^2,
\end{equation}
using a scale length, $L$, of 2.5 kpc and a characteristic scale height, $\beta$, of 200 pc. We assume that the Solar System is located 8.5 kpc from the Galactic Center and falls along the x=0, z=0 plane. Finally, using the integrated $f_{\rm GW}$ and $e$ combined with randomly chosen positions in the Galaxy, we calculate the $h_c$ and corresponding SNR for detection by LISA for each of these 2100 systems using Equation \ref{eq:h_c}. Note that during this procedure and throughout this study, we assume all NSs have a mass of 1.4 \Msun. 

Green points in Figure \ref{fig:lisa_sensitivity} show the subset of those 2100 systems that will have an SNR$>$7 when observed by LISA for a four-year mission. In this particular Milky Way realization, we find 256 DNSs. Many of these may be confused with double white dwarfs. However, binaries that evolve in frequency space over the lifetime of the LISA mission, $T_{\rm LISA}$, will have measurable chirp masses, $\mathcal{M}_c$, allowing the heavier DNSs ($\mathcal{M}_c=1.22\Msun$ for two 1.4 \Msun\ NSs) to be differentiated from their lower-mass white dwarf analogs ($\mathcal{M}_c=0.52\Msun$ for two 0.6 \Msun\ white dwarfs) \citep{kyutoku19}. The limiting frequency allowing this measurement can be determined \citep{nelemans01}:
\begin{equation}
    f_{\rm chirp} \geq 1.75\times 10^{-3} \left( \frac{\mathcal{M}_c}{1.22\ \Msun} \right)^{-5/11} \left( \frac{T_{\rm LISA}}{4\ {\rm yr}} \right)^{-6/11} {\rm Hz}.
\end{equation}

The vertical dotted line in Figure \ref{fig:lisa_sensitivity}, which shows this limit on $f_{\rm GW}$ for a four-year LISA mission, demonstrates that a subset of these DNSs ought to have measurable chirp masses. Given LISA's sensitivity, those DNSs with $f_{\rm GW}>f_{\rm chirp}$ will be easily identifiable as being comprised of two NSs \citep{seto19}. On the other hand, DNSs with $f_{\rm GW}<f_{\rm chirp}$ can be confused with Galactic double white dwarfs; from Equation \ref{eq:h_c}, $h_c\sim\mathcal{M}_c^{5/3}f^{2/3}d^{-1}$, and a binary comprised of two 0.6 \Msun\ WDs will need to be $\approx$4 times closer than an analogous DNS system with the same $h$ and $f_{\rm GW}$. Since most DNSs will be found within the Galactic Plane at $\sim$10 kpc, the intrinsic faintness of WDs at distances larger than a few hundred pc, the poor position determination on the sky by LISA and confusion in the densely packed Galactic Plane all combine to make it unlikely that optical follow-up will be able to rule out a double WD scenario for these systems.

In addition to the green points in Figure \ref{fig:lisa_sensitivity} representing a single Milky Way realization, we generate $10^5$ separate DNSs using the same procedure except with random merger times within the last 100 Myr. The grey contours in this Figure represent the overall distribution of DNSs in this plane, expected from a constant merger rate.


We run 100 separate Milky Way realizations to determine the statistical distribution of the number of expected Milky Way DNSs observable by LISA. We fit the number of DNS detections to a Gaussian distribution, finding the best fit mean of the distribution from our 100 Milky Way realizations:
\begin{equation}
    N_{\rm DNS} =
    \begin{cases}
    240 \left( \frac{R_{\rm MW}}{210\ {\rm Myr}^{-1}} \right), & {\rm MW:\ 4-year\ mission}\\
    330 \left( \frac{R_{\rm MW}}{210\ {\rm Myr}^{-1}} \right), & {\rm MW:\ 8-year\ mission}.
    \end{cases}
    \label{eq:LISA_rate_MW}
\end{equation}
For a more pessimistic rate estimate of 42 Myr$^{-1}$, as derived by the Milky Way population of DNSs \citep{pol19}, we find LISA detection rates of 46 (64) for a 4-year (8-year) LISA mission. Since these rates are determined from random sampling, uncertainties on $N_{\rm DNS}$ within the Milky Way scale with $\sqrt{N_{\rm DNS}}$. Roughly 25\% of the DNSs detected in the MW by LISA will have measurable chirp masses.

Based on the current estimate of the Milky Way DNS merger rate, we therefore conclude that LISA will almost certainly observe a handful of DNSs. Since these are Poisson processes, the rates in Equation \ref{eq:LISA_rate_MW} can be scaled up and down arbitrarily as rate estimates improve with future observations and analysis. Although we opt to not include it here, one can trivially propagate this Poisson distribution with uncertainties on the DNS merger rate.


\subsection{DNSs in Other Nearby Galaxies}

What about the nearby M31 and M81 galaxies? Returning to Figure \ref{fig:lisa_sensitivity}, we see that systems in these galaxies are detectable by LISA for a much smaller time, since these systems have $h_c$ above the LISA sensitivity curve only once $f_{\rm GW}\gtrsim10^{-2.5}$, corresponding to a merger time of $\sim 10^5$ yr. Even with the optimal orbital $f_{\rm GW}$, systems within M81 typically do not have an SNR above 2. On the other hand, Figure \ref{fig:lisa_sensitivity} shows that systems within Andromeda may produce detectable $h_c$ \citep[see also][]{seto19}. Using the same rate of DNS mergers in Andromeda as in the Milky Way (this is likely an underestimate, as Andromeda is somewhat more massive than the Milky Way), we repeat the procedure used for the Milky Way, generating 100 random realizations of M31. Setting a limit of SNR$>$7, we find:
\begin{equation}
    N_{\rm DNS} =
    \begin{cases}
    1.2 \left( \frac{R_{\rm M31}}{210\ {\rm Myr}^{-1}} \right), & {\rm M31:\ 4-year\ mission}\\
    4.3 \left( \frac{R_{\rm M31}}{210\ {\rm Myr}^{-1}} \right), & {\rm M31:\ 8-year\ mission}.
    \end{cases}
    \label{eq:LISA_rate_M31}
\end{equation}
Uncertainty on these rates follow a Poisson distribution, and therefore can be scaled up and down arbitrarily as estimates on the DNS merger rate in M31 are refined. Note that the DNS nature of these sources will be immediately apparent from $\mathcal{M}_c$, since these systems will all be ``chirping'' and furthermore the distance to Andromeda is well-determined.

The detectability of DNSs in M31 has been recently discussed by \citet{seto19}, who find $\approx$5 systems ought to be identified within a 10-year LISA mission. This author uses a DNS merger rate estimate of 500 Myr$^{-1}$, $\approx$2.5 times larger than the estimate we use. However this author also uses a SNR of 10 for their detection threshold rather than our limit of SNR$>$7. Therefore, the expected number of DNSs we find here are consistent with those found by \citet{seto19}.




\begin{figure*}
	\includegraphics[width=0.99\textwidth]{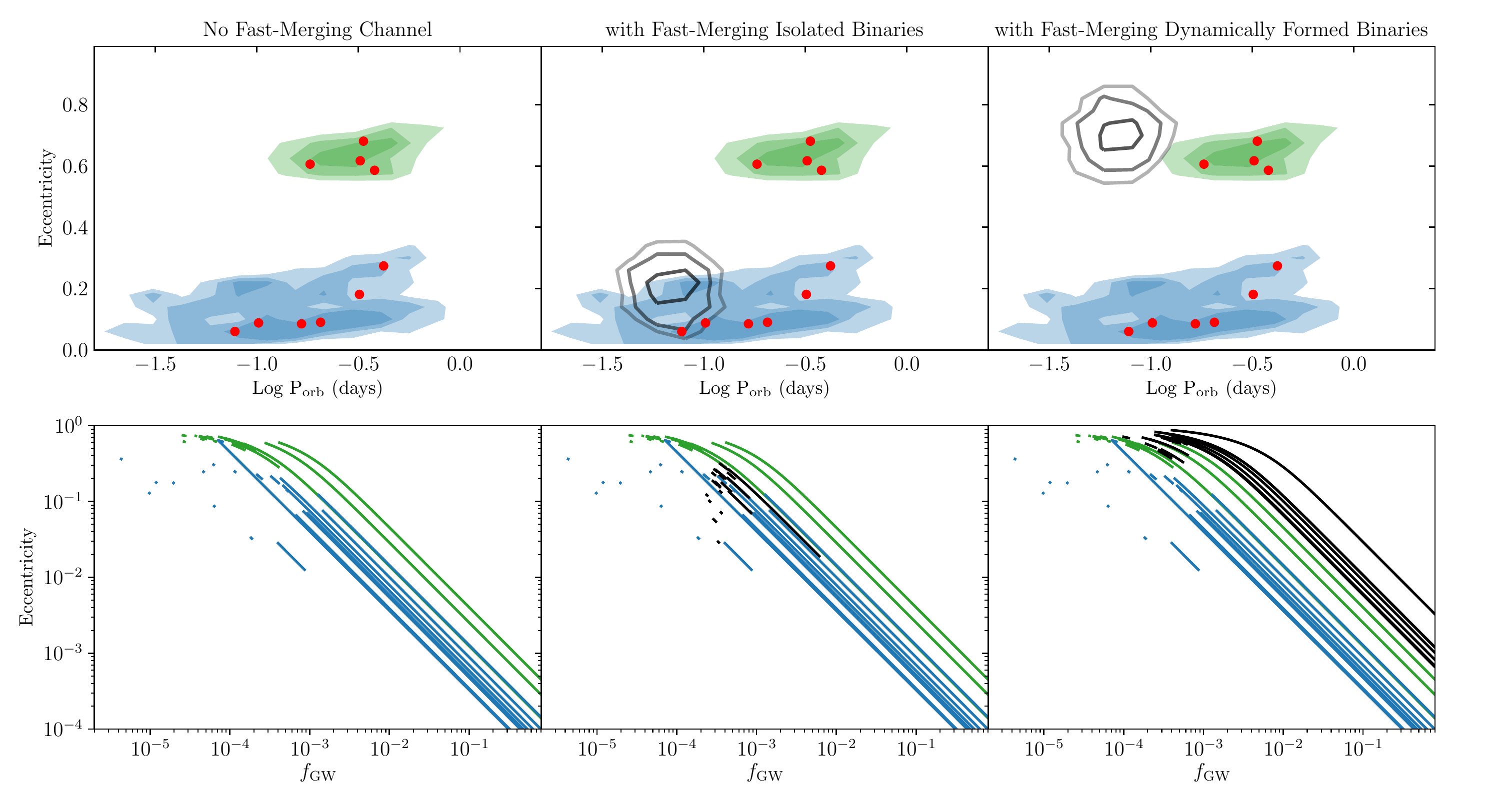}
    \caption{{\bf Top Row:} $\Porb-e$ distributions for three separate models for the formation of DNSs. Blue and green distributions show representative models for the formation of the low- and high-eccentricity DNSs observed in the Milky Way (red points). Black contours in the second and third panels show two different ad hoc distributions for a putative fast-merging DNS evolutionary channel. {\bf Bottom Row:} The evolution of DNSs in $f_{\rm GW}-e$ over as they circularize and inspiral due to GR. Depending on  the characteristics of the fast-merging channel, DNSs will evolve along separate tracks in $f_{\rm GW}-e$ space.}
    \label{fig:Porb_ecc}
\end{figure*}

\section{A Fast-Merging Channel?}
\label{sec:results}

Several studies have argued that if merging DNSs are responsible for the nucleosynthesis of $r$-process material in the Universe, a fast-merging channel that creates, evolves, and merges DNS within $\sim$10 Myr is required \citep[e.g.,][]{komiya14, matteucci14,safarzadeh17,Safarzadeh19}. More recent studies were able to reproduce the enrichment of Milky Way stars with a delay time as long as $\sim$100 Myr \citep[see discussion in][]{vangioni16}. However, the discovery of $r$-process enrichment in two ultra-faint dwarf galaxies, Reticulum II \citep{ji16a} and Tucana III \citep{hansen17}, suggests that a fast-merging channel is again required to form DNS mergers early enough so that a second generation of stars can incorporate the merger products \citep{beniamini16a, safarzadeh18}. \citet{zevin19} have recently invoked similar arguments to explain the $r$-process enrichment observed in many globular clusters.

What might cause these different evolutionary channels? One possibility deals with the formation of DNSs through isolated binary evolution including a phase of Case BB mass transfer, in which a NS accretor enters a second mass transfer phase when its stripped helium star companion evolves into a giant star \citep{delgado81}. The most up-to-date simulations predict that this Case BB mass transfer phase ought to be stable, forming DNSs with orbital periods as short as $\approx$1 hour \citep{tauris13, tauris15}. Other simulations have shown that, for certain combinations of parameters, this phase of Case BB mass transfer may be unstable \citep{dewi03, ivanova03}, in which case DNSs could form with even shorter merger times. Since the ultra-stripped star emits little mass upon core collapse, such systems are expected to form with very low eccentricities \citep{tauris15}.

Another option was proposed by \citet{andrews19b}, who suggest that the high eccentricity subpopulation of DNSs in Figure \ref{fig:DNS_evolution} is consistent with being formed dynamically in globular clusters then kicked out into the field. Indeed, B2127$+$11C is a member of the globular cluster M15 and has parameters consistent with the high-eccentricity, short-orbital period DNSs in the field (see Figure \ref{fig:DNS_evolution}). Initial population studies of globular clusters suggested that LISA may be sensitive to dynamically formed compact object binaries, including, based on crude scaling estimates, tens of double neutron stars \citep{benacquista99, benacquista01}. Later, more detailed globular cluster models found only a few dynamically formed DNSs would merge within a Hubble time \citep{grindlay06, ivanova08, lee10, belczynski18, ye19a, ye19b} and may produce of order one system detectable by LISA \citep{kremer18a}. Nevertheless, the similarity of B2127$+$11C with other DNSs in the field suggests that the dynamical formation scenario may still be relevant \citep{andrews19b}. Since dynamical formation tends to produce systems with eccentricities drawn from a thermal distribution \citep{heggie75}, this putative fast-merging channel ought to have a much higher eccentricity distribution than those formed through case BB mass transfer in isolated binaries.


Despite the circularizing effects of GR, most systems will maintain a residual eccentricity as they evolve through the LISA band, with a value depending on the exact scenario forming a putative fast-merging channel. Those DNSs with only upper limits on $e$ necessarily formed with low eccentricities.
LISA's ability to detect eccentricities in binary orbits as small as 10$^{-2}$ affords a unique opportunity to discern between various evolutionary channels forming DNSs, analogous to what has already been shown for double white dwarfs \citep[e.g.,][]{willems07} and double black holes \citep[e.g.,][]{breivik2016,nishizawa17, SamsingDOrazio:2018, DOrazioSamsing:2018}.  

To quantitatively test the eccentricity distributions for different DNS formation scenarios, we use toy models for DNS formation. We first include functional models for the formation through isolated binary evolution of the observed low- and high-eccentricity DNSs in the Milky Way. \citet{andrews19b} show that these separate populations can be reasonably modeled through isolated binary evolution, by randomly generating systems immediately prior to the second SN, then dynamically evolving them through core collapse \citep[see also,][]{andrews19a}. The low-eccentricity systems are modeled with a circular pre-SN orbit, with a log-normal orbital separation distribution ($\mu=0.2$, $\sigma=0.4$, in units of \Rsun), and an isotropic SN kick of 50 km s$^{-1}$. We reproduce the high-eccentricity DNSs in a similar way, but using a log-normal pre-SN orbital separation distribution ($\mu=0$, $\sigma=0.2$, in units of \Rsun) and a SN kick velocity of 25 km s$^{-1}$. These models for low- and high-eccentricity DNSs \citep[which are adapted from][]{andrews19b} immediately after the SN are shown in the top row of panels in Figure \ref{fig:Porb_ecc} as blue and green contours, respectively. 

In the first column of panels in Figure \ref{fig:Porb_ecc}, we show only the two functional models to reproduce the observed low- and high-eccentricity DNSs in the Milky Way. The next two columns of panels additionally contain toy models for the two putative fast-merging formation scenarios (black contours). Detailed simulations of both isolated binary evolution through Case BB mass transfer and dynamical formation within globular clusters are outside of the scope of this Letter. For these two toy models, we randomly generate DNSs with a log-normal distribution in orbital separation ($\mu=-0.1$, $\sigma=0.2$ in units of \Rsun) and a normal distribution in eccentricity ($\mu=$0.2 and 0.7 for the two models, both with $\sigma=0.1$). 
In the bottom row of panels, we show how the eccentricities of these systems decrease as they evolve through the LISA band.

For each of these four populations (low-eccentricity DNSs, high-eccentricity DNSs, and the two ad hoc models for a fast-merging channel) we model the evolution through the LISA band following a similar procedure used to calculate the $h_c$ distribution of DNSs shown in Figure \ref{fig:lisa_sensitivity}: we generate 10$^5$ separate systems for each population (so they are well sampled), evolve them forward for a randomly drawn time corresponding to the time until merger (selected from a uniform distribution within the past 100 Myr) due to GR, place them in a random position in the Milky Way, and calculate the SNR of a LISA detection. We record the system's eccentricity if it produces an SNR$>$7 within LISA. 

\begin{figure}
	\includegraphics[width=0.99\columnwidth]{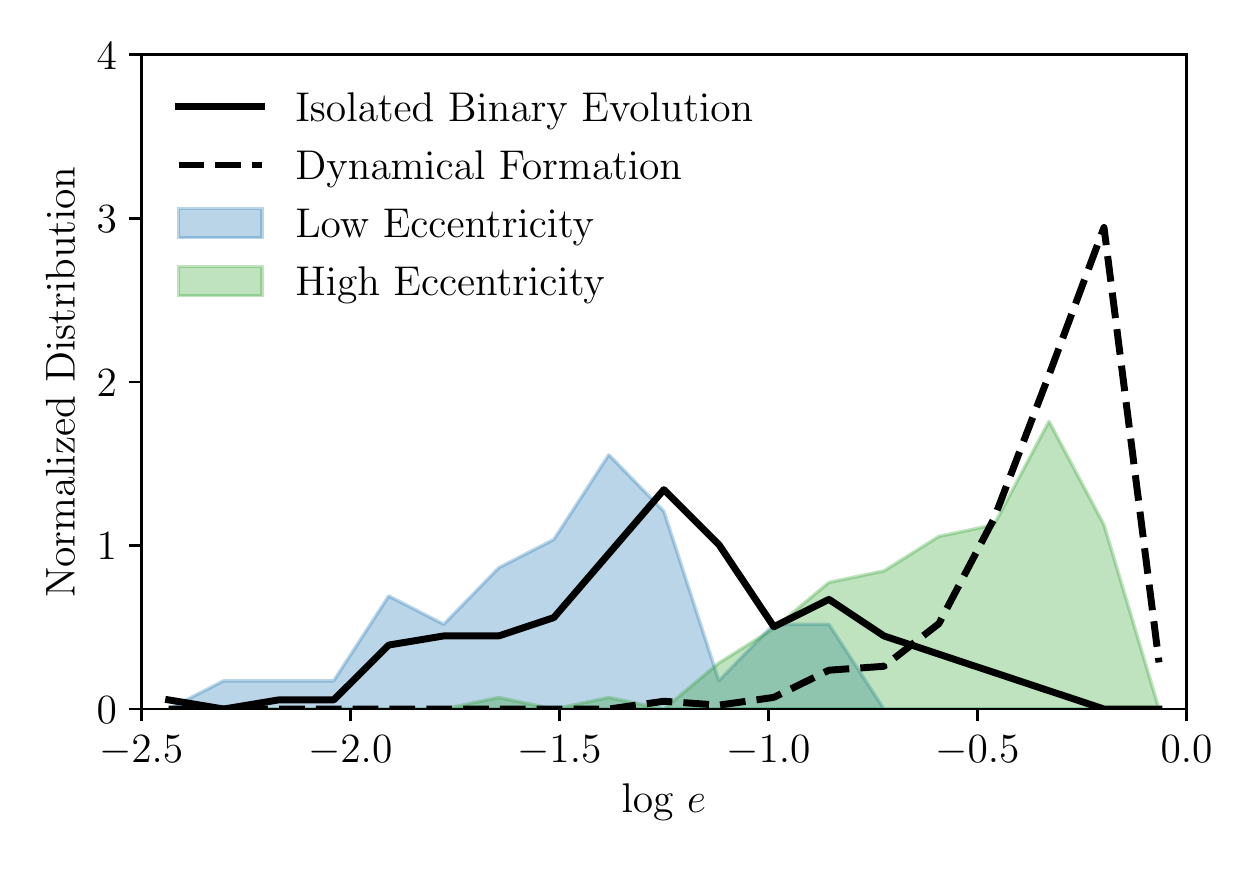}
    \caption{Eccentricities of DNSs identified within the LISA band with SNR$>$7. Blue and green distributions are designed to match the low and high eccentricity DNSs in the Milky Way, respectively. The two black distributions (different line styles) demonstrate the eccentricities expected from two ad hoc models for putative fast-merging DNSs.}
    \label{fig:ecc}
\end{figure}

Figure \ref{fig:ecc} shows the resulting normalized eccentricity distributions. Whereas our models for the Galactic DNSs produce broad eccentricity distributions, depending on the particular model chosen, the eccentricities of DNSs within the LISA band vary substantially. If a fast-merging channel produces a detectable contribution to the overall DNS merger rate, Equation \ref{eq:delta_e} indicates that an eccentricity precision of 10$^{-2}$ is sufficient to discern between the different DNS formation scenarios.

\section{Discussion \& Conclusions}
\label{sec:conclusions}

Using the Milky Way DNS merger rate of 210 Myr$^{-1}$ derived from LIGO \citep{ligo_GWTC1}, we find that a 4-year (8-year) LISA mission will detect on average 240 (330) DNSs. Approximately 25\% of those will be ``chirping,'' allowing for their characterization as DNSs through their chirp mass. The remaining $\approx$75\% will likely be indistinguishable from lower-mass double white dwarfs. 

Using a more pessimistic rate of 42 Myr$^{-1}$ based on the Galactic population of DNSs, we find LISA will detect 46 (65) DNSs for a 4-year (8-year) mission. However, the census of Milky Way DNSs is incomplete since Doppler shifting of the radio waves that pulsars emit makes detecting DNSs with orbital periods $\lesssim$1 hour extremely challenging. If these systems exist in significant numbers, the nature of gravitational wave orbital decay implies that they were formed with similarly short orbital periods. The presence of such fast-merging DNSs can help resolve the difference between the DNS merger rates as determined by LIGO and the galactic population of DNSs. 
Since LISA bridges the orbital period gap separating DNSs detected with radio waves and by LIGO, it can detect DNSs that form with short orbital periods. Furthermore, with its ability to measure orbital eccentricities as small as 10$^{-2}$, LISA affords, perhaps, the only opportunity to measure and characterize the orbits of these binaries, discerning between the different formation scenarios.

The existence of such a fast-merging channel would have profound implications on the presence of $r$-process elements in the Universe. Recent studies have shown that, if $r$-process enrichment seen in the ultra-faint dwarf galaxies Reticulum II and Tucana III is formed from the merger of DNSs, then a fast-merging channel is required \citep{beniamini16a, safarzadeh18}. Similar arguments are required to explain the existence of $r$-process enrichment in globular clusters \citep{zevin19}. 

These studies, which use binary population synthesis, suggest that such fast-merging DNSs could be formed through unstable Case BB mass transfer \citep{delgado81}, in which a NS accretor enters a second common envelope when its stripped helium star companion evolves into a giant star. Dynamical formation provides an alternative evolutionary scenario that can explain the differences between DNS merger rate estimates; DNSs in globular clusters suffer from very different selection effects \citep{bagchi11} that are not taken into account by rate estimates based on the Milky Way field DNS merger rate \citep{phinney91b}. 

Eccentricities allow for the best opportunity to discern between different formation scenarios. \citet{lau19} has recently demonstrated that LISA observations of DNS eccentricities can identify whether this Case BB mass transfer phase is stable or unstable. Here, we show that a toy model for dynamically formed DNSs will have typical eccentricities of $\gtrsim$0.3 when detected by LISA. While a detailed analysis of individual formation scenarios is outside the scope of this work, it is clear that the joint measurement of $f_{\rm GW}$ and $e$ by LISA for even a handful of DNSs provides an important diagnostic of DNS formation. 




\section*{Acknowledgements}

The authors are grateful for useful conversations with Kyle Kremer on calculating LISA signal-to-noise ratios for eccentric binaries. This work was initiated and performed in part at the Aspen Center for Physics, which is supported by National Science Foundation grant PHY-1607611. We additionally thank Josiah Schwab for organizing the conference and workshop entitled ``The Beginnings and End of Double White Dwarfs'' where much of this work took place. J.J.A.\ acknowledges support by the Danish National Research Foundation (DNRF132). K.B.\ acknowledges support from the Jeffery L. Bishop Fellowship. D.J.D. acknowledges support provided by NASA through Einstein Postdoctoral Fellowship award number PF6-170151 and funding from the Institute for Theory and Computation Fellowship.

\bibliographystyle{aasjournal}

\bibliography{references}

\end{document}